\documentclass[num-refs]{wiley-article}

\usepackage{siunitx}
\usepackage{booktabs}
\usepackage{multirow}
\usepackage{array}
\usepackage{float}
\usepackage{chngcntr}
\usepackage[table]{xcolor}
\usepackage{tabularx}
\papertype{Original Article}

\title{Safety-Driven Response Adaptive Randomisation: An Application in Non-inferiority Oncology Trials}

\author[1]{Maria Vittoria Chiaruttini PhD}
\author[2]{Lukas Pin PhD}
\author[2]{Sofía S. Villar PhD}

\affil[1]{Department of Cardiac, Thoracic, Vascular Sciences and Public Health, University of Padua, Italy}
\affil[2]{MRC Biostatistics Unit, University of Cambridge, UK }

\corraddress{Maria Vittoria Chiaruttini}
\corremail{mariavittoria.chiaruttini@studenti.unipd.it}

\fundinginfo{The authors acknowledge funding and support from the UK Medical Research Council (grants MC UU 00002/15 and MC UU 00040/03 (SSV), as well as an MRC Biostatistics Unit Core Studentship (LP), the Cusanuswerk e.V. (LP), and the PhD scholarship from the University of Padua (MVC).  SSV is part of PhaseV’s advisory board.}

\runningauthor{Chiaruttini M.V. et al.}

\begin{document}

\begin{frontmatter}
\maketitle

\begin{abstract}
The majority of response-adaptive randomisation (RAR) designs in the literature rely on efficacy data to guide dynamic patient allocation. However, their applicability becomes limited in settings where efficacy outcomes, such as survival, are observed with a random delay. To address this limitation, we introduce SAFER, a novel RAR design that leverages early-emerging safety data to inform treatment allocation decisions, particularly in oncology trials. The design is broadly applicable to contexts where prioritizing the arm with a superior safety is desirable. This is especially relevant in non-inferiority trials, to demonstrate that an experimental treatment is not inferior to the standard of care, while potentially offering improved tolerability. In such trials, an unavoidable trade-off arises: maintaining statistical efficiency for the efficacy hypothesis while integrating safety-driven adaptations through RAR. The SAFER design addresses this trade-off by dynamically adjusting the allocation proportion based on the observed association between safety and efficacy endpoints. We illustrate the performance of SAFER through a simulation study inspired by the CAPP-IT Phase III oncology trial. Results show that SAFER preserves statistical power, reduces the adverse event rate, and offers flexible adaptation speed depending on the temporal alignment of the endpoints.

\keywords{FDA Project Optimus, Efficacy, Toxicity, Adverse events, Endpoints association, Patient-benefit}
\end{abstract}
\end{frontmatter}

\section{Introduction}\label{sec1}
In contemporary drug development, optimal dosing has become a central focus, aiming for a precise, patient-centred approach in oncology. Initiatives like the Food And Drag Administration (FDA) Project Optimus \cite{FDA2024doseoptimization} underscore this shift, emphasizing the best balance between drug's efficacy and safety to maximize patient benefit and minimise toxicity.
Traditionally, early-phase clinical trials primarily prioritise safety, often focusing on major adverse events, with efficacy being largely exploratory. As development progresses, later phases are primarily driven by efficacy decisions. Although comprehensive safety data are collected, this information is typically not formally integrated into design decisions, despite the goal of improving the overall benefit-risk profile and enabling patients to remain on treatment longer.

This becomes particularly relevant in the context of the modern oncology treatment landscape, which has been transformed by the advent of targeted therapies, monoclonal antibodies, and immune checkpoint inhibitors. 
In fact, their administration typically spans prolonged treatment cycles and is often accompanied by chronic, low-grade adverse events (AEs), including fatigue, gastrointestinal disturbances, and dermatologic reactions. These AEs, although not classified as dose-limiting in the short windows of phase I trials, may significantly impair patient adherence, quality of life, and long-term tolerability in the later phases \cite{castellanos2015making, fourie2022improving}. 
Moreover, novel agents, designed to modulate specific molecular pathways, often exhibit plateauing efficacy at doses well below the maximum tolerated dose (MTD)~\cite{maritaz2022immune, PIN2024107567}, underscoring the need for a revised perspective on the interplay between safety and efficacy outcomes.
A pertinent example to illustrate this is \textit{sotorasib (LUMAKRAS)}, a KRAS G12C inhibitor approved for non-small cell lung cancer \cite{singh2025retrofit}. For this treatment a 960 mg dose was selected for pivotal trials. Subsequent analyses showed that a 240 mg dose offered similar efficacy with a better safety profile, leading the FDA to mandate a post-marketing dose optimization study. 

The above rationale motivates our interest in the following crucial question: how can a late phase adaptive design help identify alternative dosing strategies such as lower or fractionated doses, different schedules or regiments deliver, with comparable or superior efficacy and fewer side effects? This is particularly relevant to address the potential impact on phase II and III studies from dosing decisions from early trials often adopted on immediate and severe safety data. This is especially the case for agents with delayed efficacy signals and cumulative toxicity profiles that influence long-term adherence and outcomes. 

Our work aims to address the research question presented above
by leveraging Response-Adaptive Randomisation (RAR). RAR refers to a class of statistical methodologies used in clinical trial design to dynamically adjust the probability of assigning patients to different treatment arms based on accumulating outcome data \cite{hu2006theory}. By increasingly favouring arms showing early superior performance, RAR can enhance both patient outcomes and the statistical efficiency of trials \cite{sverdlov2014efficient}. This adaptability is particularly advantageous when outcomes are available shortly after treatment initiation, allowing real-time modification of allocation proportions. However, in trials where primary efficacy outcomes are delayed, such as overall survival in oncology, RAR becomes more challenging due to the unavailability of timely, informative data during enrolment. Several strategies have been proposed to address this, including the use of early surrogate endpoints that correlate with long-term outcomes \cite{huang2009using, kim2014outcome, gao2024response}. This paper introduces a RAR design that, alternatively, uses early-emerging safety data to inform treatment allocation decisions in oncology trials.

However, a critical component of a safety-driven RAR design is the consideration of the potential interplay between safety and efficacy endpoints. For instance, while some drugs may maintain or even improve efficacy after changes to the treatment regimen or dose reductions, others may show a decrease in therapeutic benefit. 
From a clinical standpoint, assigning more patients to the better-tolerated arms is not always optimal. In fact, there is an inherent risk in preferably assigning patients to a treatment that, although better tolerated, may be less effective. On the other hand, from a statistical standpoint, RAR designs may introduce challenges in preserving statistical power, especially under certain conditions. Although equal allocation does not universally guarantee maximal power in superiority trials, as highlighted by recent work by Pin et al. (2024)~\cite{pin2024response}, it remains an effective and commonly adopted strategy in many trial settings. On the contrary, in the context of non-inferiority trials with binary (Bernoulli) or time-to-event (exponential) outcomes, a balanced 1:1 allocation ratio is typically optimal due to the mathematical structure of the variance, which is a function of the mean. Under the alternative hypothesis in such trials, the non-inferiority assumption postulates that the average treatment effects across the intervention and control arms are clinically equivalent, within a narrowly defined margin of non-inferiority. This margin is usually small~\cite{guidance2010guidance}, and as a result, the outcome variances in both arms are expected to be similar. Given this, allocation strategies that aim to minimize the variance of the estimated treatment effect, such as Neyman allocation~\cite{Neyman1934,Robbins1952}, which assigns subjects in proportion to the standard deviation of the outcome in each arm, tend to approximate or even coincide with balanced allocation in these settings. Thus, attempts to skew allocation dynamically based on accumulating data, as done in RAR designs, may deviate from this optimal variance-reducing strategy.
Consequently, an unavoidable and well-established trade-off arises for such designs: to balance the goals of preserving the inferential efficiency for the primary non-inferiority outcome while incorporating safety considerations into the randomisation process through RAR. 

To this regards, Baldi Antognini et al. (2020) \cite{baldi2020compound} introduced a compound optimal allocation strategy in survival endpoint trials to address the trade-off between "ethical" and inferential concerns in the choice of the randomisation proportion. Although their work is aligned with our rationale, it is limited to single-outcome efficacy trials since it does not capture the complexity of a joint safety-efficacy decisions. 

Therefore, we propose an innovative Safety-Aware Flexible Elastic Randomization (SAFER) design. SAFER dynamically adjusts patient allocation proportions based on accumulating safety endpoint data, with the aim of leveraging the association between efficacy and safety outcomes to balance clinical benefit and statistical validity. This adaptive strategy is particularly suitable for late-phase clinical trials, as it enhances patients' treatment experience while preserving statistical power.

The remainder of this paper is structured as follows. Section~2 introduces the key components of the SAFER design: (i) the efficacy endpoint, including the associated power function and sample size calculation; (ii) the definition of the safety endpoint; (iii) the target allocation proportion; and (iv) the integration of efficacy and safety data to guide the targeted allocation proportion. Section~3 introduces the CAPP-IT study as a motivating example and presents its re-design in a non-inferiority framework using the SAFER approach. Specifically, we outline the rationale and report the results from several distinct simulation scenarios, aimed at evaluating the performance of the proposed design under varying conditions. Section~4 discusses the simulation findings, explores broader implications of the SAFER design, and concludes with a summary of key contributions and directions for future research.

\section{SAFER Design Components}\label{sec2}

\subsection{Efficacy Endpoint}

Our proposed design is specifically tailored to accommodate for the typical primary endpoint used in Phase II/III oncology trials. Specifically, we consider a time-to-event measure, such as Progression-Free Survival (PFS). Consequently, both the sample size and the primary analysis of our design will be expressed throughout this paper in terms of this designated endpoint. The sample size will be determined accordingly to the expected effect size to ensure adequate statistical power for that main primary analysis. We will now formally outline this rationale, including the underlying assumptions and notation. The exponential distribution was assumed for simulation purposes; the implications and limitations of this assumption are further discussed in Section \ref{sec4}.

Let $j=E,C$ denote the experimental treatment and control arms, respectively. The patients' primary efficacy endpoint (e) i.e., their survival time ($Y^{(e)}$ ) is assumed to be independent and identically distributed (i.i.d.) with probability density function (p.d.f) when treatment $j$ is assigned:
\begin{equation}
f(y_j^{(e)}; \theta_j^{(e)}) = (\theta_j^{(e)-1})\exp\left(-\frac{y_j^{(e)}}{\theta_j^{(e)}}\right)
\end{equation}
where $\theta_j^{(e)-1} = \lambda_j^{(e)} \in \mathbb{R}^+$ denotes the event rate. Therefore, $\theta_j^{(e)}$ represents the mean of the treatment effect, and $\theta_j^{(e)2}$ represents the variance of the treatment effect.
We focus on the case in which the study is testing for the non-inferiority of the experimental with respect to the control arm. The null hypothesis for the mean difference can be expressed as \cite{food2016non, machin2018sample}
\begin{equation}
    H_0:\theta_C^{(e)}-\theta_E^{(e)}\ge m  \quad H_1:\theta_C^{(e)}-\theta_E^{(e)} < m,
\end{equation}
where $m$ in the non-inferiority margin. Based on this hypothesis, and under the assumptions that the means are strictly positive, and the standard regularity conditions hold (to ensure the asymptotic normality of the Maximum Likelihood Estimations), we can construct a Wald statistic as follows \cite{baldi2020compound}:
\begin{equation}
W = \frac{\hat{\theta}_C^{(e)} - \hat{\theta}_E^{(e)} - m}
{\sqrt{Var(\hat\theta_C^{(e)} - \hat\theta_E^{(e)})}}
\sim \mathcal{N}(0,1),
\end{equation}
where the variance is estimated by

\begin{equation}
\sqrt{Var(\hat\theta_C^{(e)} - \hat\theta_E^{(e)}) }=\sqrt{N^{-1}\cdot (\hat{\theta}_C^{(e)2} + \hat{\theta}_E^{(e)2})},
\end{equation}

with $N$ equal to the sample size. Although both the semi-parametric Cox model and the parametric exponential survival models are consistent estimators under exponentiality, in practice, the semi-parametric model often converges more quickly and accurately in simulations \cite{kalbfleisch2002statistical, hosmer2008applied}. This is primarily due to its reliance on partial likelihood, fewer distributional assumptions, and better numerical conditioning in the presence of non-informative censoring and finite samples. Therefore, we estimate the Wald statistic through the Cox model \cite{therneau2000cox}.

In practice, the Wald statistic is constructed directly on the log-hazard ratio scale. 
Let $\hat\beta$ denote the Cox model estimate of the log-hazard ratio, so that 
$\exp(\hat\beta)$ is the estimated hazard ratio (HR). Under the non-inferiority 
hypothesis expressed on the HR scale, the null value corresponds to a pre-specified 
margin $\text{HR}_0$ (e.g., $\text{HR}_0 = 1.25$). The test statistic is therefore

\begin{equation}
W = \frac{\hat\beta - \log(\text{HR}_0)}
{SE(\hat\beta)}
=
\frac{\log(\exp(\hat\beta)) - \log(\text{HR}_0)}
{SE(\hat\beta)},
\end{equation}

where $SE(\hat\beta)$ is the standard error obtained from the Cox partial likelihood. 
This is a standard Wald test, since it standardizes the difference between the 
maximum partial likelihood estimator $\hat\beta$ and its null value by its estimated 
standard error. Under regularity conditions and large samples, $\hat\beta$ is 
approximately normally distributed, implying that $W \sim \mathcal{N}(0,1)$ under 
$H_0$. The same construction is used at both interim and final analyses.
We also incorporate an early stopping feature into our designs, recognizing that this feature, when combined with RAR, can offer advantages over RAR alone \cite{jennison2023comment}. For simplicity, we have focused on only two analyses: an interim analysis (k=1) and the final analysis (k=2). Therefore, we will provide test statistics \( W_{k=1} \) and \( W_{k=2} \), estimated based on the first group available at the interim analysis and the complete sample at the end of follow-up, respectively.

To estimate the minimum number of observed progression events $P$ needed for the study, we construct a $H_0$ rejection region as $R=\left\{ -\infty, z_{1 - \beta} \right\}$, 
with 
\begin{equation}
z_{1 - \beta}= \sqrt{P\cdot(\log(HR_{H_1})-m)^2 \pi_E \cdot (1 - \pi_E)}- z_{1 - \alpha}, 
\label{equation:5}
\end{equation}
where $\alpha$ is the significance level, $\beta$ is the type-II Error rate, $HR_{H_1}$ is the hazard ratio ($\lambda_E/\lambda_C$) under the alternative hypothesis, and $\pi_E$ is the allocation proportion to the experimental arm (E).  
Therefore, given the rejection region, we can calculate the minimum number of events $P$ needed to test the non-inferiority on the primary endpoint between treatments, based on the following power function definition \cite{wassmer2016group}.
\begin{equation}
Power = P_{H_1}(Z_1 \in \mathcal{R}_1) + P_{H_1}(Z_1 \in \mathcal{C}_1, Z_2 \in \mathcal{R}_2)
\label{equation:6}
\end{equation}
In equation~\eqref{equation:6}, $Z_1$ and $Z_2$ are the statistic tests at the interim and final analyses. $P_{H_1}(Z_1 \in \mathcal{R}_1)$ is $power_1$ (interim analysis power) i.e., the probability at the interim, under the alternative hypothesis ($H_1$), to fall into $R_1$, the region of rejection of the null hypothesis ($H_0$). Moreover, $P_{H_1}(Z_1 \in \mathcal{C}_1, Z_2 \in \mathcal{R}_2)$ is $power_2$ (final analysis power) i.e., the probability, under $H_1$ of reject $H_0$ at the final analysis given that the study continued at interim. Therefore, $power_2$ is given by a bivariate integral where $R_2$ and $C_1$ ($ 1-R_1$) indicate the rejection regions of $H_0$ and continuation at the final and interim analyses, respectively.  

After the calculation of the minimum number of events $P$ to reach an overall power of at least 0.8  (assuming $\alpha$ = 0.05,  $\pi_E$ = 0.5, the $H_1$ and the $m$ as defined below in Section \ref{sec3}), we adjust the total number of events needed due to the application of the Lan-DeMets O’Brien-Fleming $\alpha$-spending function \cite{jennison1999group} to account for $k=2$ analyses. 
Finally, due to the survival primary endpoint, the overall sample size ($N$) is estimated by dividing the total number of events needed ($P$) by the probability to observe at least one event during the follow-up, i.e., $Pr(\text{at least one event in time } t) = 1 - e^{-\lambda\cdot t}$ where lambda is the weighted (across arms) average event rate expected in the sample and t the maximum follow-up time.

\subsection{Safety Endpoint}
The safety endpoint in our design is the time to dose-reduction/drug discontinuation due to low drug tolerability or AEs.
The patients' safety endpoint (s) (\( Y^{(s)} \)), is assumed to be i.i.d.,
with p.d.f. when treatment $j$ is assigned: 
\begin{equation}
f(y_j^{(s)}; \theta_j^{(s)}) = \theta_j^{(s)-1} \exp\left(-\frac{y_j^{(s)}}{\theta_j^{(s)}}\right)
\end{equation}
where \( \theta_j^{(s)} \) is the average time to dose reduction/drug discontinuation in arm \( j \). 
The safety model and its estimation will directly inform the RAR component of the design, as detailed in the next subsection.

\subsection{Allocation Proportion}
For our proposed design, the Neyman allocation \cite{Neyman1934, Robbins1952, hu2006theory} is a reasonable choice of allocation proportion to target. First, the Neyman allocation,
maximizes the statistical power of a widely used Wald test to detect the true difference between arms; second, under the assumption of exponentially distributed survival endpoint, the Neyman proportion will consistently assign the majority of patients to the treatment that patients are observed to discontinue later \cite{baldi2020compound} (or tolerate for the longest). Note that the latter is not always true for other endpoint distributions, such as the binary case \cite{pin2025revisiting}, where the Neyman allocation can favour the worst treatment to minimize the variability of the Wald test.
In our design, the Neyman target allocation proportion is estimated based on the estimated safety average time to dose reduction as indicated below:

\begin{equation}
\hat\pi_E =\frac{\hat\theta_E^{(s)}}{\hat\theta_E^{(s)} + \hat\theta_C^{(s)}}
\label{equation:8}
\end{equation}

It is worth highlighting that we define $\pi_E$ as the \emph{true} target allocation proportion, computed by plugging the true values of $\theta_E^{(s)}$ and $\theta_C^{(s)}$ into equation \ref{equation:8}, in accordance with the simulation settings used for safety data generation. In contrast, $\hat{\pi}_E$ denotes the estimated target allocation proportion, obtained by substituting $\hat{\theta}_E^{(s)}$ and $\hat{\theta}_C^{(s)}$ i.e., the estimations of the respective parameters, into the same equation.

Specifically, in our simulation design, we allow for \( u = 1, \ldots, U \) update times for $\hat{\pi}_E$, which occur independently of the $k$-analyses defined by the group-sequential design (where interim efficacy analyses are assumed to occur only once, at $k=1$).
Therefore, the randomisation proportion \( \hat{\pi}_E \) is updated sequentially:
\begin{itemize}
    \item For patients enrolled before the first update time ($u=1$), \( \hat{\pi}_E \) is set to 0.5. This is the {\it burn-in} stage of the RAR feature of this design \cite{thorlund2018key, Tang2025}.  
\item For patients enrolled between update time $u=1$ and $u=2$, \( \hat{\pi}_E \) uses the estimate of  $\hat\theta_E^{(s)}$ and  $\hat\theta_C^{(s)}$ from the first update time.
\item This pattern continues, with patients enrolled between $u=2$ and $u=3$ using the estimate of  $\hat\theta_E^{(s)}$ and  $\hat\theta_C^{(s)}$ from the second update time, and so on, until the final update at $u=U$.
 
\end{itemize}

The number of update times ($U$) and the corresponding calendar occasions should be specified in the study protocol. The updates may be equally spaced across the enrolment period or not, depending on the design objectives and practical considerations. 
We fixed equally spaced updates every three months, starting from month 3 through month 45, assuming a total enrolment period of 48 months. Crucially, each  \( \hat{\pi}_E \) is estimated using all available data up to a specific update time. 

\subsubsection{Efficacy and Safety Association}
Because the allocation strategy is primarily driven by safety considerations, there is an inherent risk of preferentially assigning patients to a treatment that, although better tolerated, may be less effective. To address this concern, our design incorporates a safeguard mechanism that exploits the estimated relationship between efficacy and safety outcomes. Consequently, the design can detect situations in which an apparent safety advantage is offset by insufficient efficacy, thereby avoiding allocation to a futile or inadequately active treatment arm.
A key feature of the SAFER design is its ability to adapt allocation probabilities based on joint evidence from both endpoints. Unequal allocation among arms is permitted only when improved tolerability is observed and this safety advantage is not undermined by weak efficacy results. If the efficacy data fail to demonstrate benefit, the randomization probability reverts toward a 1:1 ratio.
To quantify how strongly improved safety should influence allocation in light of efficacy evidence, we introduce a weight parameter, $\hat{\Phi}$, which ranges from 0 to 1. It is defined as the cumulative distribution function (CDF) of the $W$-value (on the Z-scale) associated with the estimated treatment effect on PFS, obtained from the chosen estimation model.
For instance, if a Cox proportional hazards model is used, a negative estimated coefficient (and corresponding $z$-value) indicates a reduction in risk for the experimental arm. Therefore, to map $\hat{\Phi}$ onto the $[0,1]$ scale in a consistent direction, the sign of the estimated $z$-value must be inverted. Conversely, when using a parametric model in which a positive $z$-value reflects favourable efficacy for the experimental arm, the sign is left unchanged.
For simplicity, in the following formulas we refer only to $\hat{\Phi}$, assuming that the $z$-value has already been transformed according to the modelling framework so that larger values consistently correspond to stronger evidence of efficacy in favour of the experimental treatment.

Although the W-statistic may be unstable in very early stages of a trial when the number of PFS events is limited, the SAFER design is primarily conceived for confirmatory settings (e.g., Phase IIb or Phase III trials), where event accumulation occurs more predictably and treatment effect estimates progressively stabilize. In practice, RAR procedures are typically implemented after an initial burn-in period with equal allocation, ensuring that adaptation occurs only once a sufficient amount of information has accrued. Moreover, a minimum information threshold (for example, a pre-specified number of PFS events) may be required before allowing $\hat{\Phi}$ to influence allocation probabilities. Under such a rule, randomization would remain 1:1 until adequate information is available, thereby mitigating the impact of early variability in the $z$-statistic. Consistent with this conservative principle, whenever the transformed $z$-value is less than or equal to zero (i.e., $\hat{\Phi} \leq 0.5$), the allocation probability reverts to complete randomization.

The choice of this weight parameter is motivated by the fact that a simple correlation metric between safety and efficacy is not appropriate in this context, as it fails to account for the specific characteristics of survival endpoints. In particular, both endpoints are subject to truncation: the time-to-event safety endpoint (e.g., time to dose reduction or drug discontinuation) and the PFS endpoint are both affected by administrative censoring. Moreover, the safety endpoint is inherently truncated by the occurrence of PFS events, since once progression occurs, follow-up ends. As a result, the two endpoints are structurally correlated, and conventional correlation coefficients may yield misleading estimates of their association.

Specifically, consider that the parameter $\hat{\Phi} = CDF(\hat{W})$ is incorporated into the allocation function as follows:
\begin{itemize}
    \item When the experimental arm is better tolerated (i.e., the estimated Neyman allocation is greater than 0.5) and $\hat{\Phi}$ exceeds a predefined threshold, the allocation probability for the experimental arm is increased.
    \item When $\hat{\Phi}$ is below the threshold, suggesting insufficient evidence of efficacy, allocation probabilities revert toward equal randomization.
\end{itemize}

Since $\hat{\Phi}$ ranges from 0 to 1, the default threshold is set at 0.5 (corresponding to a $z$-statistic equal to zero). However, sensitivity analyses should be conducted to assess the impact of alternative threshold choices.
The \textit{SAFER} function is defined as:
\begin{equation}
\text{SAFER}(\hat{\pi}_{E}) =
\begin{cases} 
    0.5 & \text{if } \hat{\Phi} \leq 0.5 \\ 
    0.5 + (\hat{\pi}_{E} - 0.5) \cdot \left( 1 - \left( 1 - \dfrac{\hat{\Phi} - 0.5}{0.5} \right)^\eta \right) 
    & \text{if } 0.5 < \hat{\Phi} < 1 \\ 
    \hat{\pi}_{E} & \text{if } \hat{\Phi} = 1
\end{cases}
\label{equation:10}
\end{equation}

\begin{figure*}[t]
\centering\includegraphics[width=\textwidth]{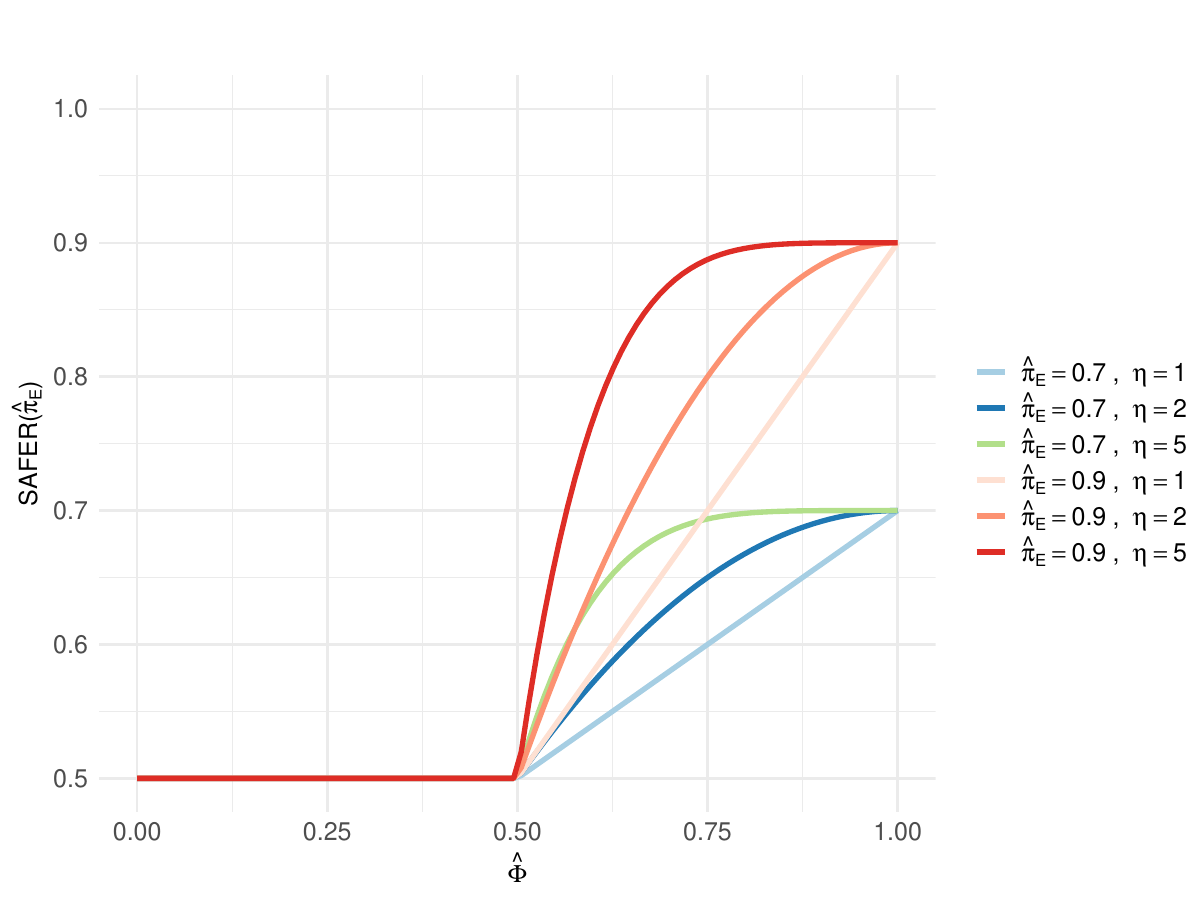} 
\caption{Target function of the SAFER design for two target allocations, $\hat{\pi}_E = 0.7$ and $\hat{\pi}_E = 0.9$, each evaluated under three shape parameters $\eta = 1, 2, 5$. The SAFER function increases monotonically in $\hat\Phi$ and reflects more aggressive allocation as both $\hat{\pi}_E$ and $\eta$ increase.
}
\label{fig:1}
\end{figure*}

Note that the maximum allocation proportion used in the SAFER design corresponds to the estimated Neyman allocation $\hat{\pi}_{E}$ when $\hat{\Phi} = 1$, while the minimum is fixed at 0.5, representing complete randomisation, when $\hat{\Phi} \leq 0.5$. Sensitivity analyses should also explore scenarios with upper bounds below 1 (e.g., 0.9) to assess robustness.
Moreover, the parameter $\eta$ is the elastic parameter which controls the curvature of the function: \( \eta = 1 \) corresponds to a linear function, while \( \eta > 1 \) results in a concave shape, as illustrated in Figure~\ref{fig:1}. Although convex shapes (\( \eta < 1 \)) are theoretically allowed, we did not explore the values of \( \eta < 1 \), as the linear function is already considered a conservative choice to adjust the allocation proportion in cases of small $\hat\Phi$ values.

In practical applications, the choice of the elastic parameter $\eta$ should be guided by simulation studies tailored to the specific clinical setting. Such simulations should evaluate operating characteristics including type I error control, statistical power, allocation proportions under various safety–efficacy scenarios, and the risk of allocating patients to a potentially futile arm. In this sense, $\eta$ should be viewed as a calibration parameter that governs the aggressiveness of the adaptation, rather than as an arbitrary constant. The final choice should reflect a balance between ethical considerations related to patient allocation and the statistical properties required for confirmatory inference.

\section{SAFER Design Evaluation}\label{sec3}

This section details the extensive simulation studies conducted to systematically assess the performance of the proposed SAFER design methodology. These simulations aim to evaluate key operating characteristics including statistical power and the estimated allocation proportion to the experimental arm under various scenarios. The findings presented herein are based on a specific case study, which we redesign using SAFER and help us provide a comprehensive understanding of the design features and its potential advantages in practical trial settings.

\subsection{Motivating example: a follow-up to the CAPP-IT study}
The motivating study is a multicentre, randomised, double-blind, placebo-controlled phase III trial included in the NIHR portfolio and conducted across 10 UK sites (CAPP-IT study; \textit{British Journal of Cancer}, 2012, \textbf{107}, 585--587~\cite{corrie2012randomised}). Patients with colorectal or breast cancer receiving palliative treatment with single-agent capecitabine were randomised to receive either concomitant pyridoxine or a matching placebo. The primary objective was to assess whether pyridoxine could reduce the need for capecitabine dose reduction and thereby improve patient outcomes. The safety endpoint was defined as the incidence of capecitabine dose reduction, while the efficacy endpoint was PFS.
The original sample size calculation, based on the safety endpoint, required 270 patients to detect a reduction in dose reduction incidence from 30\% to 15\% with 80\% power, accounting for potential dropouts. However, recruitment was slower than anticipated, and the study was closed prematurely after enrolling only 106 patients. Pyridoxine did not demonstrate a statistically significant effect on the objective response rate (odds ratio: 1.37; 95\% CI: 0.475--3.96). Notably, there was a result suggesting decreased PFS with pyridoxine: median PFS was 7.4 months in the pyridoxine group and 9.9 months in the placebo group (hazard ratio: 1.62; 95\% CI: 0.91--2.88).
In the following subsections, we present a re-design of this study using our SAFER approach. Specifically, we formulate a non-inferiority hypothesis on the primary efficacy endpoint (PFS), while for the safety-driven RAR, we focus on the time to dose reduction or drug discontinuation due to capecitabine-induced side effects as the safety endpoint. Moreover, we assume a maximum duration of the trial equal to 60 months, considering a uniformly distributed enrolment of 48 months and 12 months of follow-up. Assuming a monthly event rate in the control arm (\(\lambda_C\)) of 0.069, which corresponds to a median PFS of approximately 10 months, and a $HR_{H0}$ of 1.25 ($\lambda_E/\lambda_C$), the median PFS in the experimental arm must exceed 8 months to demonstrate non-inferiority with respect to the control arm.

\subsection{Modelling the association between safety and efficacy endpoints}
\begin{figure*}[t]
	\centering
	\includegraphics[width=\textwidth]{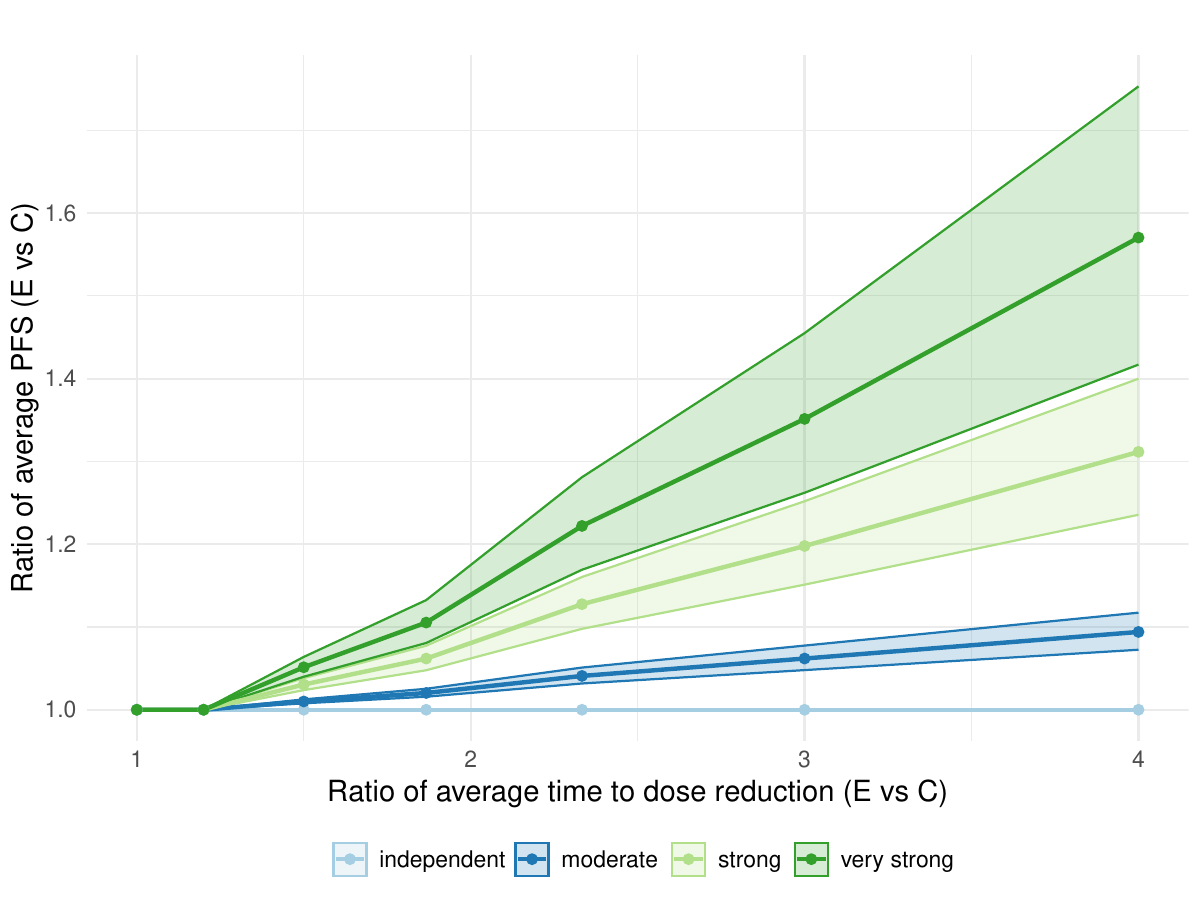} 
	\caption{Relationship between the ratio of average time-to-dose reduction/drug discontinuation and the ratio of average PFS in experimental (E) vs control (C) arms, at different level of endpoints association: independent, moderate, strong, very strong). Bootstrap (10,000 iterations) 95\% confidence intervals.}
	\label{fig:2}
\end{figure*}

This section details how the relationship between endpoints is incorporated into our simulated datasets.
Specifically, we assume that completing \( \leq 3 \) drug cycles results in an average PFS 
which aligns with the expected PFS in the control group (since the average number of completed cycles in the control group is 3). Moreover, we assume that the number of "extra" (\( >3 \)) completed drug cycles without AEs has an impact on PFS such that the average PFS for each patient \( i \) is defined by \( \mu_i = e^{\gamma_0 + \gamma_1 \cdot (\text{extra-cycles})} \), where \( e^{\gamma_0} = 434.78 \) is the average PFS in days in the control group, and \( \gamma_1\) is the estimated effect on PFS due to each new completed "extra" drug cycle.  
The parameter \( \gamma_1 \) is modelled as a gamma-distributed random variable, with the expected value $E[\gamma_1]$ set at 0.001, 0.005, 0.01, 0.03, and 0.05. The shape parameter is fixed at 50 (arbitrary choice), while the scale parameter is computed as the ratio of the expected value to the shape parameter (i.e., \(\text{scale} = \text{expected value} / 50\)). The endpoints link function is shown in Figure~\ref{fig:2}. After obtaining the mean for each patient \( i \), we calculated the corresponding daily rate 
\( \lambda_i \) as \( 1 / \mu_i \), such that the \( PFS_i \sim \text{Exp}(\lambda_i) \).
This framework allow us to incorporate cases in which the expected PFS increases as the number of completed cycles increases, as proxy of enhanced drug tolerability.

\subsection{Metrics to assess performance}
We define a \emph{safety-driven RAR design} as one in which the allocation proportion is updated solely based on the estimated Neyman allocation proportion. In contrast, we refer to the \emph{SAFER design} when the allocation updates incorporate the SAFER function in equation \ref{equation:10}, which accounts for the observed association between efficacy and safety.

To compare these adaptive designs with a fixed allocation approach, we provide benchmark results focusing on statistical power. Specifically, we assess power using both a theoretical (non-simulation-based) approach as defined in equations~\eqref{equation:5} and~\eqref{equation:6} and a simulation-based approach, which employs exponentially distributed survival times as outcomes. The latter allows us to empirically assess the validity of the asymptotic assumptions underlying the theoretical approximation.
Specifically, in Figure \ref{fig:5} the theoretical power is maximized by $\pi_E=0.5$ and decreases to approximately 0.79 when $\pi_E=0.6$, 0.74 when $\pi_E=0.7$, and 0.63  when $\pi_E=0.8$. In Table \ref{tab4} we report the power at increasing allocation proportions, using exponentially distributed outcomes. The results closely align with the theoretical results in Figure \ref{fig:5}, suggesting that for further comparisons, the normal approximation can be reliably used in this setting. Interestingly, it seems that the Information Fraction (IF) at interim does not modify the observed loss of power.

Moreover, through simulations, we evaluate several distinct scenarios to assess the design under various conditions (described below), providing insights into its statistical properties and broader implications. 
As metrics, we report the overall study power and type-I-error rate, calculated as the proportion of simulated trials that rejected $H_0$ under $H_1$ and $H_0$, respectively, the $power_1$, calculated as the proportion of simulated trials that rejected $H_0$ under $H_1$ at interim analysis (probability of early efficacy stop), the estimated allocation proportion to the experimental arm, defined as the average proportion of patients assigned to treatment $E$ at the end of each simulated trial, and the AE rate per arm  calculated as the total number of AEs divided by the total person-years, i.e., the number of patients multiplied by the follow-up duration in years, to yield a rate per patient-year. 
For each scenario, we conduct 10,000 iterations, ensuring a Monte Carlo error of less than 0.5\% for both power and type-I error rate control~\cite{morris2019using}.

\subsection{Simulating scenarios}
In this section, we will describe in detail the distinct scenarios used for the simulation study.

\textbf{Scenario 0: Safety-driven RAR design (without a SAFER approach) under independence between the efficacy and safety endpoints.}

Scenario 0 is specifically designed to isolate the impact of safety-driven response-adaptive randomization (RAR) on inference for the efficacy endpoint under the assumption of independence between safety and efficacy. In this setting, adaptation in allocation is induced solely by differences in the safety endpoint, while the efficacy endpoint remains statistically independent of safety. Importantly, the SAFER down-weighting mechanism is not applied, so this scenario represents a purely safety-driven RAR design and serves as a methodological benchmark.
The median time to dose reduction or drug discontinuation is fixed at 1.5 months in the control group. In the experimental group, it is varied across 1.5, 2.25, 3.5, and 6 months, corresponding to target allocation proportions $\pi_E = 0.5$, 0.60, 0.70, and 0.80, respectively. This setup ensures that adaptation in $\hat{\pi}_E$ is driven exclusively by safety differences. We set $U = 15$, resulting in updates of $\hat{\pi}_E$ every three months until month 45, following a three-month burn-in period with 1:1 randomization.
By separating safety-driven adaptation from efficacy differences and excluding the SAFER correction, Scenario 0 evaluates whether adaptive allocation based solely on safety information distorts inference on the efficacy endpoint. Moreover, this scenario allows a direct assessment of error control under adaptive allocation. When  the experimental arm is inferior in efficacy, the rejection probability reflects potential inflation of the Type I error rate. Demonstrating appropriate error control in this setting establishes that response-adaptive randomization per se does not compromise nominal Type I error. 

\textbf{Scenario 1: Complete randomization design under diverse levels of association between the efficacy and safety endpoints.}

This scenario is designed to assess the impact of diverse levels of association between the efficacy and safety endpoints on statistical power and AEs rate. Specifically, we simulate increasing levels of association between the endpoints, with $E[\gamma_1]$ set to 0.001, 0.005, 0.01, 0.03, and 0.05, corresponding to very weak, weak, moderate, strong, and very strong associations, respectively. Moreover, we set the median AEs time in each arm corresponding to $\pi_E$ from 0.5 (no difference in safety between arms) to 0.8. To evaluate the effect of association between endpoint at the net of RAR, we simulate under a 1:1 allocation ratio (fair coin toss).

\textbf{Scenario 2: Safety-driven RAR design (without a SAFER approach) under diverse levels of association between the efficacy and safety endpoints.}

This scenario is designed as a counterpart to scenario~1 and aims to assess the impact on power, allocation proportion, and AEs rate of the safety-driven RAR design. Therefore, at each time point \( u = 1, \ldots, U \), we compute $\hat{\pi}_E = \hat{\theta}_E^{(s)} / (\hat{\theta}_E^{(s)} + \hat{\theta}_C^{(s)})$
and use it as the allocation probability to arm $E$ for the randomization of patients in the subsequent three months.
Under this scenario, we expect power to decrease in the presence of weak association between endpoints, and to increase as association strengthens, potentially due to the enhanced effectiveness of the experimental treatment on the efficacy endpoint driven by prolonged well-tolerated drug exposure.

\textbf{Scenario 3: SAFER design: impact on power under diverse levels of association between the efficacy and safety endpoints, varying the $\eta$ values.}

This scenario is designed to evaluate the impact on power, allocation proportion, and AEs rate of the SAFER design, by varying the elastic parameter \( \eta \) values. Specifically,  scenarios 3a and 3b report the results of simulations conducted with \( \eta = 1 \) and \( \eta = 5 \), respectively. These values correspond to the absence of elasticity and a high degree of elasticity in the target SAFER function, as reported in Figure \ref{fig:1}.

\textbf{Scenario 4: Impact of differential timing in the observation of endpoint data on $SAFER(\hat{\pi}_E)$.}

This scenario aims to assess the speed at which $SAFER(\hat{\pi}_E)$ reaches the target level, when the median times of the safety and efficacy endpoints are either close or far apart i.e., delayed efficacy outcome \cite{kim2014outcome}. For instance, we expect a faster pattern when the median PFS is close to the median time to dose reduction or drug discontinuation because there will be enough PFS events for the $\hat\Phi$ parameter estimation, early in the recruitment period. On the other hand, when evidence of the positive association between safety and efficacy emerges later in the recruitment, we expect the SAFER function to down-weight the allocation proportion for a longer period until sufficient evidence on PFS becomes available. Under this scenario, 
the median PFS ranges from 3 to 24 months.

\textbf{Scenario 5: Impact of informative drop-out (related to the safety endpoint) on $SAFER(\hat{\pi}_E)$ and power.}

This scenario aims to assess the impact of informative drop-out on the operating characteristics of the SAFER design.
We define informative drop-out as a situation in which a patient is observed until experiencing a side effect and subsequently drops out of the study. 
In practice, this means that the patient is followed until dose reduction or drug discontinuation occurs, after which neither the time to PFS nor the occurrence of the progression event is observed.
We simulated the average allocation proportion and power under varying informative drop-out rates, ranging from 5\% (very low) to 25\% (very high), with drop-outs uniformly distributed over the accrual period. To address the potential bias introduced by informative drop-out, we considered the strategies for handling intercurrent events as defined in the Estimands framework \cite{kahan2024estimands}. Specifically, we considered the \textit{composite strategy}, where the time-to-PFS is defined as the last available follow-up time, with the occurrence of the progression = yes.

\textbf{Scenario 6: Impact of under-reported safety events on $SAFER(\hat{\pi}_E)$ and power.}

This scenario aims to assess the impact of under-reported safety events on the operating characteristics of the SAFER design. 
Unlike the informative drop-out phenomenon—where patients are observed until an adverse event occurs and then drop out—in the case of under-reported safety events, it is not recorded at all and only the survival outcome (i.e., progression and the corresponding time-to-PFS) is available \cite{seruga2016under}.

\subsection{Simulation Results}

The total sample size required to achieve at least a power value of 0.8 is 881 (about 18 patients enrolled per months). However, due to the interim analysis, the total number of events needed increases to 888 when IF = 0.5. 

Table~\ref{tab1} presents the results of the simulations under Scenario~0 (independence between endpoints). Both the estimated allocation proportion of patients assigned to the experimental arm $N_E/N$ and the power closely align with the true target allocation proportion $\pi_E$ and the theoretical power reported in Table \ref{tab4}, respectively. Minor discrepancies can be attributed to the burn-in period, as Table \ref{tab4} reports results based on unequal allocation without any burn-in, whereas Scenario~0 includes this initial adjustment phase (about 60 patients). The type-I error rate remains close to the nominal value of 0.05, with minor increases. However, these increases do not show a consistent pattern and are likely due to simulation variability. Therefore, neither the safety-driven RAR design nor the information fraction (IF) appears to negatively affect type-I error control under the null hypothesis.

\begin{table}[ht]
\centering
\caption{Estimated allocation proportion, power, and type-I error rate summary: Scenario 0}
\label{tab1}
\begin{tabular*}{\textwidth}{@{\extracolsep{\fill}} >{\centering\arraybackslash}p{1.5cm} >{\centering\arraybackslash}p{1.5cm} >{\centering\arraybackslash}p{3cm} >{\centering\arraybackslash}p{1.5cm} >{\centering\arraybackslash}p{1.5cm} >{\centering\arraybackslash}p{1.5cm} >{\centering\arraybackslash}p{1.5cm}}
\toprule
\multicolumn{2}{c}{\textbf{Allocation}} & \textbf{Power/Type-I error rate} & \multicolumn{4}{c}{\textbf{Information Fraction}} \\
$\pi_E$ & $N_E/N$ & & 0.2 & 0.3 & 0.4 & 0.5 \\
\midrule
\multirow[t]{2}{*}{0.5} 
& 0.5 & Power             & 0.80  & 0.80  & 0.80  & 0.80 \\
& 0.5   & Type-I error rate & 0.048 & 0.050 & 0.048 & 0.053 \\
\midrule
\multirow[t]{2}{*}{0.6} 
& 0.59 & Power            & 0.78  & 0.79  & 0.78  & 0.78 \\
& 0.59 & Type-I error rate & 0.047 & 0.050 & 0.055 & 0.053 \\
\midrule
\multirow[t]{2}{*}{0.7} 
& 0.69 & Power            & 0.74  & 0.74  & 0.73  & 0.73 \\
& 0.69   & Type-I error rate & 0.049 & 0.051 & 0.051 & 0.053 \\
\midrule
\multirow[t]{2}{*}{0.8} 
& 0.78 & Power            & 0.66  & 0.66  & 0.65  & 0.66 \\
& 0.78  & Type-I error rate & 0.050 & 0.055 & 0.056 & 0.051 \\
\bottomrule
\end{tabular*}
\par\vspace{1ex}
{$\pi_E$ = True target allocation proportion by simulation setting; $N_E/N$ = Proportion of patients assigned to the experimental arm using a safety-driven RAR design.}
\end{table}

Table~\ref{tab2} presents the results of the simulations under Scenarios~1, 2, 3a, and 3b. Overall, it shows how the SAFER method effectively balances the benefits of a safety-driven RAR design with the need to maintain statistical validity.
Scenario~1 is simulated using a complete randomization approach. Therefore, the observed increase in both $power_1$ and overall power can be attributed to the influence of the safety endpoint on the efficacy endpoint. Specifically, the longer the time to AEs in arm~E, the longer the PFS tends to be in the same arm, which increases the probability of rejecting the null hypothesis. This trend is amplified by the strength of the association between the endpoints, with $power_1$ reaching 0.82 when $\pi_E = 0.8$ and the endpoint association is very strong.

Scenario~2 is simulated using a safety-driven RAR design, without the SAFER approach. As the estimated allocation proportion $N_E/N$ increases, both $power_1$ and the overall power decrease in the presence of very weak or weak associations between endpoints. Conversely, when the strength of association increases, a gain in power is observed. This improvement is driven by the meaningful positive effect on PFS resulting from drug exposure free from AEs, as described in Section~3.2.

Under Scenario~3a, the SAFER design is implemented with the elastic parameter $\eta$ fixed at 1. In the presence of weak endpoint associations, the SAFER approach effectively shrinks the estimated allocation proportion towards 0.5, thereby restoring statistical power while still favouring the better-tolerated arm though to a lesser extent. Conversely, when the association between endpoints is strong, the design allows higher allocation proportions in favour of the experimental arm without compromising statistical power.
A similar pattern is observed in Scenario~3b, where the higher value of the elastic parameter $\eta=5$ allows for a more aggressive approach in reaching the target allocation proportion. This comes at the cost of a slight reduction in power compared to Scenario~3a.

A crucial metric for evaluating the benefit of the proposed approach is the reduction in the AE rate per patient-year. Specifically, under the CR design (Scenario~1), the AE rate decreases as $\pi_E$ increases, from 1.28 when $\pi_E = 0.5$ to approximately 1.05 when $\pi_E = 0.8$. However, thanks to the safety-driven RAR design in Scenario~2 and the SAFER design in Scenarios~3a and 3b, the probability of being assigned to the better-tolerated arm increases. As a result, a larger reduction in the AE rate per patient-year is observed, with values dropping below 1 for high $N_E/N$ values and strong endpoint associations.

\begin{table}[ht]
\centering
\caption{Estimated allocation proportion, power, and AEs rate summary: Scenarios 1, 2, 3a, and 3b (IF = 0.5)}
\label{tab2}
\resizebox{\textwidth}{!}{%
\begin{tabular}{cccccccccccccccccc}
\toprule
\toprule
\multicolumn{2}{c}{\textbf{}} 
& \multicolumn{4}{c}{\textbf{Complete randomization}} 
& \multicolumn{4}{c}{\textbf{Safety-driven RAR}} 
& \multicolumn{4}{c}{\textbf{SAFER design $(\eta=1)$}} 
& \multicolumn{4}{c}{\textbf{SAFER design $(\eta=5)$}} \\
$\pi_E$ & Association & $N_E/N$ & $p_1$ & $p$ & AEr & $N_E/N$ & $p_1$ & $p$ & AEr & $N_E/N$ & $p_1$ & $p$ & AEr & $N_E/N$ & $p_1$ & $p$ & AEr \\
\midrule
\multirow[t]{5}{*}{0.5} 
& Very Weak & 0.50 & 0.21 & 0.80 & 1.28 & 0.50 & 0.21 & 0.80 & 1.28 & 0.50 & 0.21 & 0.80 & 1.28 & 0.50 & 0.21 & 0.80 & 1.28 \\
& Weak      & 0.50 & 0.21 & 0.80 & 1.28 & 0.50 & 0.21 & 0.80 & 1.28 & 0.50 & 0.21 & 0.80 & 1.28 & 0.50 & 0.21 & 0.80 & 1.28 \\
& Moderate  & 0.50 & 0.21 & 0.80 & 1.28 & 0.50 & 0.21 & 0.80 & 1.28 & 0.50 & 0.21 & 0.80 & 1.28 & 0.50 & 0.22 & 0.80 & 1.28 \\
& Strong    & 0.50 & 0.21 & 0.80 & 1.27 & 0.50 & 0.20 & 0.78 & 1.27 & 0.50 & 0.21 & 0.79 & 1.27 & 0.50 & 0.21 & 0.79 & 1.27 \\
& Very Strong & 0.50 & 0.21 & 0.80 & 1.27 & 0.50 & 0.20 & 0.79 & 1.27 & 0.50 & 0.21 & 0.79 & 1.27 & 0.50 & 0.21 & 0.79 & 1.27 \\
\midrule
\multirow[t]{5}{*}{0.6} 
& Very Weak & 0.50 & 0.21 & 0.80 & 1.23 & 0.59 & 0.22 & 0.79 & 1.22 & 0.52 & 0.22 & 0.81 & 1.23 & 0.54 & 0.22 & 0.80 & 1.23 \\
& Weak      & 0.50 & 0.23 & 0.80 & 1.23 & 0.59 & 0.22 & 0.80 & 1.22 & 0.53 & 0.22 & 0.81 & 1.23 & 0.54 & 0.22 & 0.82 & 1.23 \\
& Moderate  & 0.50 & 0.23 & 0.83 & 1.23 & 0.59 & 0.24 & 0.82 & 1.22 & 0.53 & 0.25 & 0.84 & 1.23 & 0.54 & 0.25 & 0.84 & 1.23 \\
& Strong    & 0.50 & 0.28 & 0.87 & 1.23 & 0.60 & 0.28 & 0.86 & 1.22 & 0.53 & 0.29 & 0.87 & 1.23 & 0.55 & 0.29 & 0.87 & 1.22 \\
& Very Strong & 0.50 & 0.33 & 0.91 & 1.23 & 0.60 & 0.31 & 0.90 & 1.22 & 0.53 & 0.33 & 0.91 & 1.22 & 0.55 & 0.32 & 0.91 & 1.22 \\
\midrule
\multirow[t]{5}{*}{0.7} 
& Very Weak & 0.50 & 0.22 & 0.81 & 1.16 & 0.69 & 0.19 & 0.75 & 1.12 & 0.55 & 0.22 & 0.80 & 1.15 & 0.58 & 0.21 & 0.81 & 1.14 \\
& Weak      & 0.50 & 0.25 & 0.84 & 1.16 & 0.69 & 0.21 & 0.78 & 1.12 & 0.55 & 0.24 & 0.84 & 1.15 & 0.59 & 0.24 & 0.84 & 1.14 \\
& Moderate  & 0.50 & 0.27 & 0.87 & 1.16 & 0.69 & 0.24 & 0.81 & 1.11 & 0.56 & 0.27 & 0.87 & 1.15 & 0.59 & 0.27 & 0.87 & 1.14 \\
& Strong    & 0.50 & 0.41 & 0.95 & 1.15 & 0.69 & 0.35 & 0.92 & 1.11 & 0.58 & 0.41 & 0.95 & 1.14 & 0.62 & 0.39 & 0.95 & 1.13 \\
& Very Strong & 0.50 & 0.55 & 0.98 & 1.15 & 0.69 & 0.49 & 0.96 & 1.11 & 0.60 & 0.54 & 0.98 & 1.13 & 0.64 & 0.53 & 0.99 & 1.12 \\
\midrule
\multirow[t]{5}{*}{0.8} 
& Very Weak & 0.50 & 0.23 & 0.81 & 1.05 & 0.78 & 0.16 & 0.66 & 0.92 & 0.57 & 0.22 & 0.81 & 1.01 & 0.62 & 0.20 & 0.81 & 0.99 \\
& Weak      & 0.50 & 0.28 & 0.87 & 1.05 & 0.78 & 0.19 & 0.73 & 0.92 & 0.59 & 0.27 & 0.87 & 1.01 & 0.64 & 0.25 & 0.86 & 0.98 \\
& Moderate  & 0.50 & 0.34 & 0.92 & 1.04 & 0.78 & 0.24 & 0.79 & 0.91 & 0.60 & 0.33 & 0.92 & 1.00 & 0.66 & 0.31 & 0.91 & 0.97 \\
& Strong    & 0.50 & 0.62 & 0.99 & 1.03 & 0.79 & 0.46 & 0.95 & 0.90 & 0.66 & 0.59 & 0.99 & 0.96 & 0.71 & 0.56 & 0.99 & 0.94 \\
& Very Strong & 0.50 & 0.82 & 1.00 & 1.03 & 0.79 & 0.66 & 0.99 & 0.89 & 0.70 & 0.80 & 1.00 & 0.93 & 0.74 & 0.76 & 1.00 & 0.91 \\
\bottomrule
\end{tabular}}
\par\vspace{1ex}
{AEr = Adverse event rate; Complete randomization = Scenario 1; Safety-driven RAR = Scenario 2; SAFER design $(\eta=1)$ = Scenario 3a; SAFER design $(\eta=5)$ = Scenario 3b;  $p_1$ = Power at interim; $p$ = Overall power; $\pi_E$ = True target allocation proportion by simulation setting; $N_E/N$ = Proportion of patients assigned to the experimental arm.}
\end{table}

\begin{figure*}[t]
	\centering
	\includegraphics[width=\textwidth]{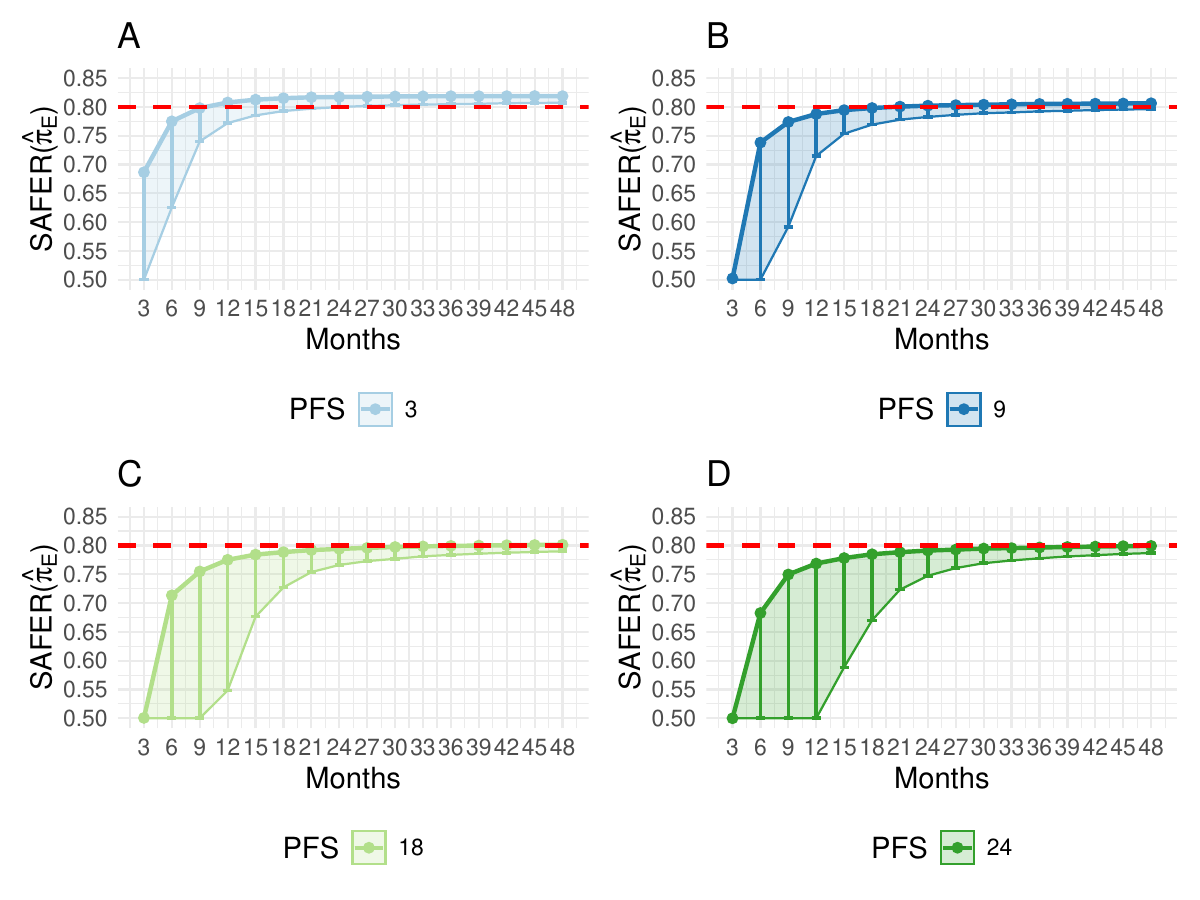} 
	\caption{$SAFER(\hat{\pi}_E)$ vs. enrolment period, under varying endpoint proximities (A: Median PFS=3 months; B: Median PFS=9 months; C: Median PFS=18 months; D: Median PFS=24 months;). True target allocation value by simulation setting=0.8; IF=0.5; $\eta$=5; Association between endpoints: very strong.}
	\label{fig:3}
\end{figure*}

\begin{figure*}[t]
	\centering
	\includegraphics[width=\textwidth]{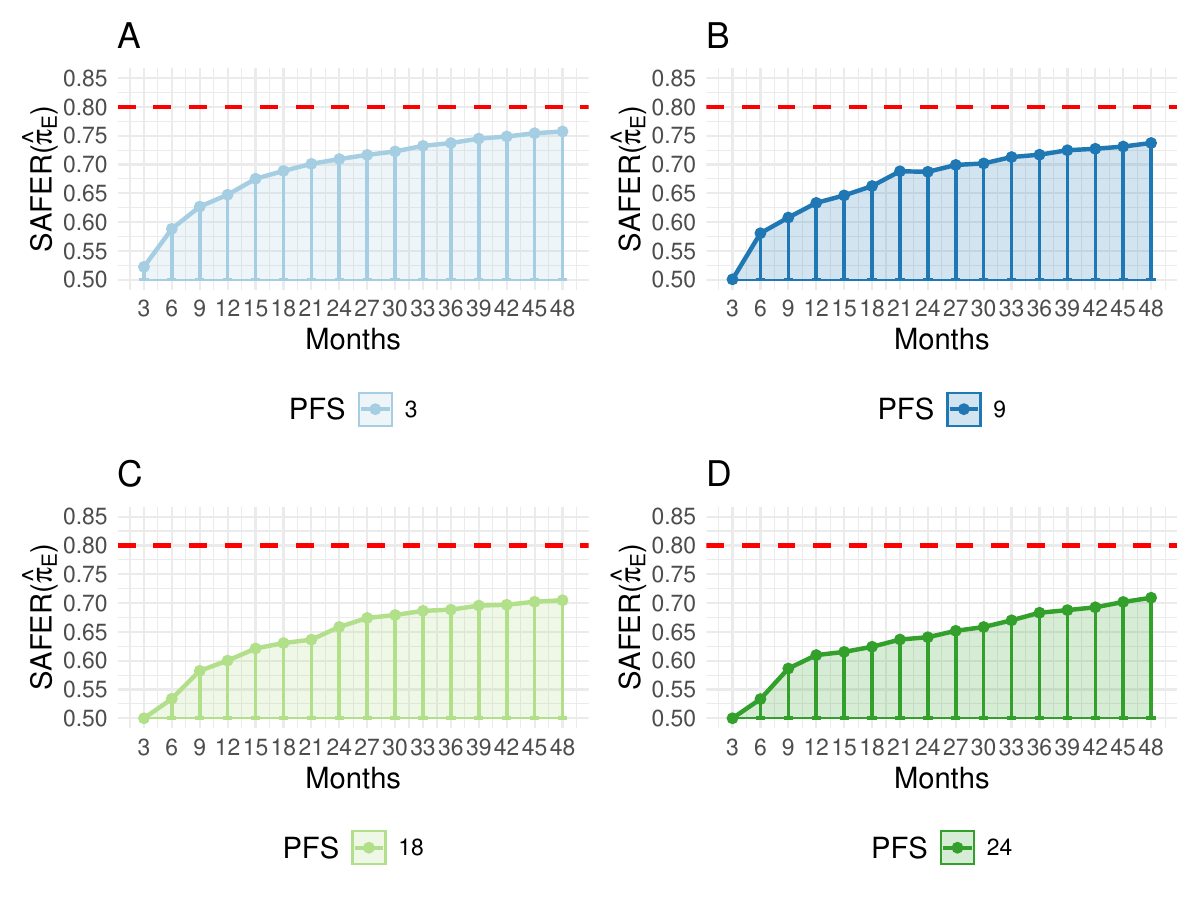} 
	\caption{$SAFER(\hat{\pi}_E)$ vs. enrolment period, under varying endpoint proximities (A: Median PFS=3 months; B: Median PFS=9 months; C: Median PFS=18 months; D: Median PFS=24 months;). True target allocation value by simulation setting=0.8; IF=0.5; $\eta$=5; Association between endpoints: weak.}
\label{fig:4}
\end{figure*}

Figure~\ref{fig:3} and Figure~\ref{fig:4} illustrate the median allocation proportion, estimated using the SAFER design, over an accrual period of 48 months. 
Through scenario 4, we underscore the critical role of endpoint timing in the proposed design. When the safety and efficacy endpoints are temporally aligned (Panels A and B), the design detects associations more rapidly, enabling faster and more confident adjustments in allocation proportions. In contrast, greater temporal separation between endpoints (Panels C and D) delays the detection of such associations and slows the adaptation process. This finding highlights a key advantage of the SAFER design: the flexibility provided by the proposed function allows for better control over the degree of adaptation, particularly in settings with delayed efficacy outcomes, a common feature in survival studies. Unlike the safety-driven RAR approach, which update the allocation proportion based solely on the safety endpoint data, the SAFER design incorporates regularization that prevents premature and extreme imbalances in allocation. This is especially relevant in non-inferiority settings, where the experimental arm could ultimately prove inferior. By moderating allocation shifts until sufficient evidence accumulates on the efficacy endpoint, the design protects patients from being preferentially assigned to an arm that performs better on safety outcomes but may fail on efficacy assessment. Importantly, this guarantees that, under the null hypothesis, the allocation mechanism does not favour an inferior treatment, even if it appears more tolerable.

Further, the study evaluated the impact of two common issues in oncology trials: informative drop-out and under-reporting. Table~\ref{tab3} reports the estimated allocation proportion $SAFER(\hat{\pi}_E)$ and the estimated statistical power at increasing percentages of drop-out and under-reporting rates. A direct comparison with the results presented in Table~\ref{tab2} for Scenario 3b with a “Very Strong” endpoint association is possible, as both the strength of the endpoint association and the parameter $\eta$ are held constant across the two settings.

The adoption of a composite strategy to address informative drop-out (Scenario 5) leads to a progressive increase in the estimated allocation proportion as the drop-out rate rises. Under this strategy, for patients who discontinue treatment prematurely, time-to-PFS is imputed as equal to the time-to-dose reduction or treatment discontinuation. As the drop-out rate increases, the proportion of patients allocated to the experimental arm ($N_E/N$) gradually converges toward the classical Neyman allocation proportion $\hat{\pi}_E$. This convergence can be explained as follows: imputing PFS using the time-to-drug reduction reduces the discrepancy between the observed PFS difference and the safety endpoint difference across treatment arms. Consequently, the down-weighting effect introduced by the SAFER approach becomes progressively attenuated. In other words, as PFS differences increasingly resemble the differences observed in time-to-drug reduction, the allocation strategy approaches the standard Neyman allocation.
In parallel, statistical power increases with higher drop-out rates. This occurs for two main reasons: i) increased number of composite events: as drop-out rises, more patients contribute events to the composite endpoint (safety plus efficacy), thereby increasing the total event count used in the analysis; ii) amplified overall treatment effect: by design, the treatment arms differ substantially with respect to the safety endpoint. As the composite endpoint incorporates this safety component, the overall effect size increases. Together, these mechanisms enhance the probability of rejecting $H_0$ under the the null hypothesis.

Although a higher drop-out rate introduces bias in the estimation process, under the hypothesis of positive association between endpoints, this bias does not necessarily compromise patient benefit. However, a critical issue arises when the experimental arm demonstrates better tolerability but worse efficacy. In such a scenario, the composite strategy may fail to adequately protect the efficacy endpoint analysis. Specifically, imputing time-to-PFS as equal to time-to-dose reduction or discontinuation may artificially inflate the apparent treatment benefit driven by safety differences, even when the experimental treatment is inferior in terms of efficacy. This may lead to substantial inflation of the type I error rate. For example, when simulating under the null hypothesis (hazard ratio for PFS equal to 2, favoring control), while maintaining a safety advantage in the experimental arm (e.g., median time-to-dose reduction/discontinuation of 6 months), the probability of rejecting the null hypothesis exceeds 80\%. For clarity, in this setting the null hypothesis refers to no efficacy benefit (or even inferiority) in PFS, despite the presence of a safety advantage in the experimental arm. The composite strategy allows the safety benefit to dominate the analysis, thereby masking the lack of efficacy and leading to erroneous rejection of the null hypothesis. Importantly, this bias is not introduced by the RAR design itself, but rather by the nature of the composite strategy—particularly when drop-out is more likely in the presence of poor tolerability, and tolerability differs significantly between arms. Therefore, when adopting a composite strategy, the treatment effect on the efficacy endpoint should be interpreted as a combined measure of both efficacy and safety since it reflects the integrated impact of tolerability on clinical outcomes. Under this interpretation only, the use of a composite strategy is reasonable and justified.

Finally, in scenario 6, as the under-report rate increases, the allocation proportion decreases. This result is expected, as the core of the proposed RAR method relies on detecting differences in the safety endpoint. The greater the under-reporting of this endpoint, the less evidence the algorithm has to adjust the allocation proportion in favour of the better-tolerated arm.

\begin{table}[ht]
\centering
\caption{Estimated allocation proportion and power summary: Scenarios 5 and 6 (IF = 0.5, $E[\gamma]=0.05$, $\eta=5$)}
\label{tab3}

\small
\setlength{\tabcolsep}{4pt}
\rowcolors{2}{gray!12}{white} 

\begin{tabularx}{\textwidth}{c l *{4}{>{\centering\arraybackslash}X}}
\toprule
\multicolumn{2}{c}{} 
& \multicolumn{2}{c}{\textbf{Informative drop-out}} 
& \multicolumn{2}{c}{\textbf{AEs under-reporting}} \\
\cmidrule(lr){3-4} \cmidrule(lr){5-6}
$\pi_E$ & Rate & $N_E/N$ & $p$ & $N_E/N$ & $p$ \\
\midrule

\multirow[t]{5}{*}{0.5} 
& Very Low    & 0.50 & 0.81 & 0.50 & 0.79 \\
& Low         & 0.50 & 0.81 & 0.50 & 0.79 \\
& Moderate    & 0.50 & 0.83 & 0.50 & 0.79 \\
& High        & 0.50 & 0.85 & 0.50 & 0.79 \\
& Very High   & 0.50 & 0.87 & 0.50 & 0.79 \\
\midrule

\multirow[t]{5}{*}{0.6}
& Very Low    & 0.55 & 0.92 & 0.55 & 0.91 \\
& Low         & 0.56 & 0.94 & 0.54 & 0.91 \\
& Moderate    & 0.56 & 0.95 & 0.54 & 0.91 \\
& High        & 0.56 & 0.96 & 0.54 & 0.90 \\
& Very High   & 0.56 & 0.97 & 0.53 & 0.91 \\
\midrule

\multirow[t]{5}{*}{0.7}
& Very Low    & 0.64 & 0.99 & 0.63 & 0.98 \\
& Low         & 0.65 & 0.99 & 0.62 & 0.98 \\
& Moderate    & 0.65 & 1.00 & 0.61 & 0.99 \\
& High        & 0.66 & 1.00 & 0.60 & 0.99 \\
& Very High   & 0.67 & 1.00 & 0.60 & 0.98 \\
\midrule

\multirow[t]{5}{*}{0.8}
& Very Low    & 0.75 & 1.00 & 0.73 & 1.00 \\
& Low         & 0.76 & 1.00 & 0.72 & 1.00 \\
& Moderate    & 0.76 & 1.00 & 0.71 & 1.00 \\
& High        & 0.77 & 1.00 & 0.70 & 1.00 \\
& Very High   & 0.77 & 1.00 & 0.69 & 1.00 \\
\bottomrule
\end{tabularx}

\normalsize

\vspace{1ex}
{\footnotesize
AE = Adverse event; $p$ = Overall power; $\pi_E$ = True target allocation by simulation setting; 
$N_E/N$ = Proportion of patients assigned to the experimental arm using a SAFER approach; 
Informative drop-out = Scenario 5; AEs under-reporting = Scenario 6.
}
\end{table}

\section{Discussion}\label{sec4}

This paper introduces a safety-driven RAR design with the aim to enhance patient-oriented research by preferentially allocating participants to treatment arms with a more favourable safety profile. This approach is particularly relevant in oncology and may support the design of new trials to assess the long-term efficacy and tolerability of novel agents that are typically administered over prolonged treatment cycles and are frequently associated with chronic and cumulative toxicities, major contributors to treatment discontinuation and reduced efficacy.
Moreover, a key feature of our design is the explicit evaluation of the interplay between efficacy and safety endpoints, an aspect increasingly emphasized by regulatory authorities and highlighted by the FDA through initiatives such as Project Optimus.

\textbf{Main Results}

Our results confirm the trade-off between safety-driven allocation design and statistical power for the efficacy endpoint analysis, in two-arm non-inferiority trials.
Therefore, we propose a \textsc{SAFER} design to provide a practical and flexible framework for incorporating safety considerations into the randomisation process while maintaining clinical and statistical rigour, by dynamically adjusting the allocation proportion based on the observed association between safety and efficacy outcomes. 
Particularly, in the presence of a negative or null association between safety and efficacy endpoints, the algorithm conservatively shrinks the allocation proportion toward 0.5 to avoid preferentially allocating patients to a safer but potentially inferior arm and to prevent loss of statistical power. Conversely, when a positive association, between safety and efficacy is detected, whether strong or weak, the algorithm allows proportionally more aggressive adaptations. In doing so, the SAFER design preserves the statistical power of the primary analysis and adhering to the hierarchical endpoints structure and the non-inferiority objective: to identify a new treatment that, given clinically equivalent efficacy, can reduce the incidence of AEs, enhance the overall patient experience, and, in certain cases, even improve drug effectiveness through a more favourable therapeutic experience due to fewer side effects. 
Notably, under associated endpoints, statistical power is not only preserved but often increases with stronger association. This can be attributed to improved efficacy outcomes resulting from longer or more tolerable exposure to the experimental treatment. 

\textbf{Considerations on design flexibility and generalizability}

The current design adopts a group-sequential framework with a single interim analysis and a final analysis. However, it is customizable, allowing for the specification of multiple interim analyses and the selection of various $\alpha$-spending functions. Additionally, a $\beta$-spending function can be incorporated to formally account for futility stopping, thereby providing greater flexibility in controlling both type-I and type-II error rates \cite{lakens2021group}.

Although the proposed design is developed within a frequentist framework, a Bayesian counterpart can be readily implemented. Specifically, the allocation proportion parameter $\hat{\pi}_E$ could be estimated using Thompson Sampling or derived from posterior probability-based decision rules \cite{thompson1933likelihood}. Similarly, the parameter $\hat\Phi$ can be interpreted as the posterior probability that the treatment effect exceeds zero. This Bayesian extension preserves the adaptive nature of the design while offering enhanced flexibility to incorporate prior information and dynamically update beliefs as data accumulate.

An important methodological consideration pertains to the generalizability of the SAFER design. While the motivating example is situated within a non-inferiority framework, the design can be customised for superiority trials. A central message of this work is to highlight the role of safety data as a guiding component in RAR designs, owing to their early availability and their potential to enhance patient-centred research. Furthermore, both the \( \arg\min(SAFER(\hat{\pi}_{E}) )\) and \( \arg\max(SAFER(\hat{\pi}_{E}) ) \), as well as the function boundaries (which, in our example, are defined as \(\text{lower} = 0.5\) and \(\text{upper} = 1\)), can be tailored to the specific requirements of the trial setting.

Our simulation results demonstrate that the proposed design, when simulated under a positive association between safety and efficacy, achieves power exceeding the nominal 80\% planned during sample size calculation. This surplus is attributable to the increased probability of rejecting the null hypothesis when the experimental treatment performs better than expected. We demonstrate that such a power surplus can enhance the likelihood of early stopping for efficacy, thereby reducing trial duration and associated costs. Alternatively, this additional power could be leveraged by incorporating further adaptive elements into the design, such as sample size re-estimation or adjustment of the non-inferiority margin.

Finally, it is important to emphasize that the clinical rationale for favouring allocation towards the arm demonstrating superior performance on the safety endpoint remains ethically and scientifically defensible, even under the assumption of independence between safety and efficacy endpoints or in the presence of an acceptable benefit–risk trade-off. In many oncology settings, improved tolerability can translate into enhanced adherence, quality of life, and overall patient experience, which are clinically meaningful outcomes in their own right.
Importantly, the potential loss of statistical power associated with unequal allocation is both predictable and quantifiable. Therefore, if necessary and operationally feasible, trial designs may incorporate a pre-specified sample size re-estimation rule at interim analyses. Such an adaptive increase in sample size, triggered when interim results are promising or when allocation imbalance reduces efficiency, would allow recovery of any power loss while preserving the ethical and patient-centred advantages of a safety-driven RAR design ~\cite{mehta2011adaptive}.

\textbf{Limitations of the parametric assumptions}

The exponential distribution is frequently adopted in methodological investigations due to its analytical tractability, but it imposes the restrictive assumption of a constant hazard function, \( \lambda(t) = \lambda \). In oncology settings, hazard rates are often time-dependent, reflecting delayed treatment effects, waning efficacy, or heterogeneous patient risk profiles. Consequently, the exponential model may provide an oversimplified representation of the underlying survival process. Importantly, inference within the SAFER framework is based on semi-parametric procedures (e.g., Cox-type estimation and log-rank–type testing), which do not require specification of the baseline hazard and therefore remain valid under general hazard structures.
A key structural feature of the exponential model is the deterministic relationship between its first two moments. If  the time \( T \sim \mathrm{Exp}(\lambda) \), then \( \mathbb{E}[T] = 1/\lambda \) and \( \mathrm{Var}(T) = 1/\lambda^{2} = \mathbb{E}[T]^2 \). As a consequence, allocation proportional to the standard deviation (Neyman allocation) is equivalent to allocation proportional to the mean time-to-event. Under this distribution, maximizing statistical efficiency for hazard ratio estimation coincides with preferentially allocating patients to the arm with the larger expected time to dose reduction. This alignment is a specific property of the exponential family and does not generally extend to more flexible survival models.
To evaluate the dependence of this alignment on the exponential assumption, we analytically examined the Weibull family, \( T \sim \mathrm{Weibull}(k,\lambda) \), with hazard function \( h(t) = k \lambda t^{k-1} \), which reduces to the exponential model when \( k = 1 \). In the Weibull case, \( \mathbb{E}[T] = \lambda^{-1/k}\Gamma(1 + 1/k) \), and the variance depends nonlinearly on both \( k \) and \( \lambda \), so the fixed quadratic relationship between mean and variance no longer holds. By varying \( (k,\lambda) \) across treatment arms, we examined configurations in which differences in expected time-to-event arise through scale effects, shape effects, or both.
Across a broad class of clinically relevant parameter configurations—particularly those corresponding to meaningful differences in expected time-to-event—we found that mean and variance move concordantly. In these settings, variance-based (Neyman) allocation continues to favor the arm with the larger mean, thereby preserving the qualitative ethical–statistical alignment underlying SAFER.
Discordance between mean and variance arises only under specific parameter combinations, typically when mean differences are negligible and dispersion differences are primarily driven by the shape parameter. Such configurations represent boundary scenarios rather than standard alternative hypotheses in confirmatory trials, which are generally powered to detect clinically meaningful differences in mean survival or hazard. When mean differences are small, any divergence between variance-driven allocation and a mean-based ethical objective is correspondingly limited.
In summary, the exponential model provides a particularly transparent benchmark in which efficiency and ethical preference coincide exactly. Nevertheless, analytic examination of the Weibull family indicates that the qualitative operating characteristics of SAFER extend beyond the constant-hazard setting and remain stable across a wide range of non-exponential survival structures.

\textbf{Association between safety and efficacy endpoints}

In the numerical illustrations, dependence between safety (time to dose reduction) and efficacy (PFS) is induced through a monotonic mechanism in which patients completing more treatment cycles are assigned a higher expected PFS. This construction provides a transparent way to generate correlation between endpoints; however, the SAFER framework itself is not conceptually restricted to a monotonic positive association.
The clinical motivation of SAFER is particularly aligned with exposure–response settings characterized by plateau behavior, as frequently observed with immunotherapies and certain targeted agents. In such cases, comparable efficacy may be achieved at lower dose intensity, so that reduced toxicity can coexist with equal (or near-equal) clinical benefit. Under this paradigm, allocating more patients to the safer arm does not compromise efficacy and may in fact prevent unnecessary exposure without loss of therapeutic effect.
More generally, three clinically relevant exposure–response scenarios can be distinguished. 
First, when improved safety is accompanied by comparable efficacy—consistent with a plateau-type dose–response relationship frequently observed with immunotherapies and targeted agents—the SAFER principle directly reinforces a favorable benefit–risk profile. In this setting, allocating more patients to the safer arm does not entail a loss of therapeutic benefit and may reduce unnecessary toxicity without compromising clinical effect.
Second, when an arm demonstrates superior safety but meaningfully lower efficacy, SAFER does not advocate adoption of an inferior treatment. The design operates within a confirmatory framework in which efficacy is formally and rigorously evaluated. Safety-driven allocation therefore functions alongside, rather than in place of, efficacy assessment. If an arm fails to demonstrate adequate efficacy, it would not satisfy the overall criteria for clinical acceptability, irrespective of its safety profile.
Thirdly, in classical cytotoxic paradigms, or in specific contexts where stronger biological or immune activation can simultaneously improve efficacy and increase adverse events, efficacy and toxicity may be positively associated. In such cases, the benefits and harms of treatment are intrinsically linked, and the ethical question becomes one of acceptable trade-offs. 
Consequently, although the illustrative data generation mechanism assumes a monotonic positive relationship between safety and efficacy, the conceptual basis of SAFER takes into account plateau effects and trade-offs between toxicity and efficacy. We recognize that more flexible joint modeling of safety and efficacy, including non-monotonic or negatively associated structures, represents an important direction for future methodological development.

\textbf{Informative drop-out and composite strategy}

Scenario 5 investigates informative drop-out handled through a composite strategy, in which time-to-PFS for patients who discontinue treatment prematurely is set equal to time-to-drug discontinuation. As the drop-out rate increases, both the estimated allocation proportion and the empirical power rise substantially, in some settings approaching 100
This pattern warrants careful interpretation. The apparent gain in power should not be interpreted as improved sensitivity for detecting a PFS effect. Rather, with increasing drop-out, the estimand progressively shifts from a pure efficacy measure toward a composite of safety (time to discontinuation) and efficacy (PFS). Consequently, the hypothesis effectively being tested differs from the original non-inferiority objective defined on PFS alone. The increasing rejection probability reflects the growing contribution of the safety component to the composite endpoint, particularly given the sizeable simulated treatment effect on tolerability.
From an analytical and regulatory standpoint, this distinction is fundamental. Under the composite construction, a treatment that is superior in safety but inferior in efficacy could still lead to rejection of the null hypothesis. In such circumstances, the type I error rate with respect to the efficacy estimand may be materially inflated. The issue therefore lies not in the SAFER response-adaptive randomization mechanism, but in the way the intercurrent event (treatment discontinuation) is incorporated into the endpoint definition. By redefining PFS through imputation at discontinuation, the analysis implicitly blends safety and efficacy effects and alters the estimand.
Accordingly, the increase in empirical power observed at higher drop-out rates should not be interpreted as evidence of design robustness, but rather as a consequence of changing the target of inference. If a composite strategy is adopted, the estimand must be explicitly defined as a joint safety–efficacy estimand, and the associated hypothesis, interpretation, and operating characteristics must be aligned with this objective.
Alternative approaches consistent with the estimand framework may offer greater conceptual clarity. When post-discontinuation efficacy data are available, a treatment policy strategy can mitigate bias arising from informative censoring by analyzing outcomes regardless of treatment discontinuation. Alternatively, when a biologically plausible and well-specified association exists between safety and efficacy, a hypothetical strategy may be considered. In this setting, the efficacy endpoint is modelled conditionally on the observed safety outcome to estimate what would have occurred in the absence of the intercurrent event \cite{gogtay2021understanding}. Such an approach may preserve estimand consistency and potentially improve inferential efficiency, provided that the underlying modelling assumptions are justified \cite{kahan2024estimands, gogtay2021understanding}.
In some clinical contexts, a safety–efficacy trade-off may be clinically meaningful and even intentionally targeted. In those cases, a composite endpoint may represent an appropriate and informative estimand. However, this objective differs from the primary aim of the present work, which focuses on non-inferiority in PFS. If a joint safety–efficacy objective is pursued, this should be explicitly stated at the design stage, and the statistical framework should be calibrated accordingly.
Further methodological investigation, including extensive simulation studies under varying degrees of association between safety and efficacy and differing drop-out mechanisms, is needed to more comprehensively evaluate the operating characteristics of alternative strategies for handling informative drop-out in adaptive oncology trials.

\textbf{Conclusions}

In conclusion, our proposed safety-driven RAR design offers a significant contribution in terms of practical tools to deliver patient-centred trials by adaptive learning from quickly observable safety data to proactively mitigate the assignment to poorly tolerated treatments without negatively affecting the power of the primary trial's objective. While highly relevant to Phase III oncology, its principles are broadly applicable. Methodological research will explore alternative distributions and non-parametric methods for small samples, with empirical validation in oncology settings essential for practical refinement. Finally, future efforts will focus on extending this framework to multi-arm trials, incorporating richer safety outcome data, including digital endpoints (for more frequent adaptation), and exploring the impact of non-monotonic relationship between safety and efficacy endpoints on the design characteristics.

\clearpage

\section*{Acknowledgements}
The authors would like to thank David Robertson, Senior Research Associate at MRC Biostatistics Unit, and Connor Fitchett, Ph.D. student at MRC Biostatistics Unit, for their helpful review and insightful suggestions that improved the clarity and presentation of this work.
The authors also acknowledge the use of large language models (LLMs) to assist with generating figures and tables, and with refining grammar and wording in the manuscript. The LLMs were not used for data analysis, interpretation, or original scientific writing. All content has been carefully reviewed and verified by the authors, who take full responsibility for the integrity and accuracy of the work.

\section*{Conflict of interest}
The authors declare no potential conflict of interests.

\section*{Supporting Information}
The code supporting the findings of this study is available upon request from the corresponding author at:

\noindent
https://github.com/MariaVittoriaChiaruttini/SAFER-design. Moreover, the following Supplementary Materials are available:
\begin{itemize}
    \item Figure \ref{fig:5}: Theoretical loss of power (resulting from equations 5 and 6 of the main Manuscript) for the primary endpoint analysis, corresponding to an Information Fractions of 50\% and allocation proportions $\pi_E$, from 0.5 to 0.8.
    \item Table \ref{tab4}: Overall power using exponentially distributed outcomes with fixed allocation proportion and varying information fraction at interim analysis.
    \item A Practical Guidance for Customizing the SAFER Design.
\end{itemize}


\clearpage

\bibliography{sample}




\clearpage

\section*{Supplementary Material}

\renewcommand{\thefigure}{S\arabic{figure}}
\renewcommand{\thetable}{S\arabic{table}}
\setcounter{figure}{0}
\setcounter{table}{0}

\begin{figure}[ht]
\centering\includegraphics[width=\textwidth]{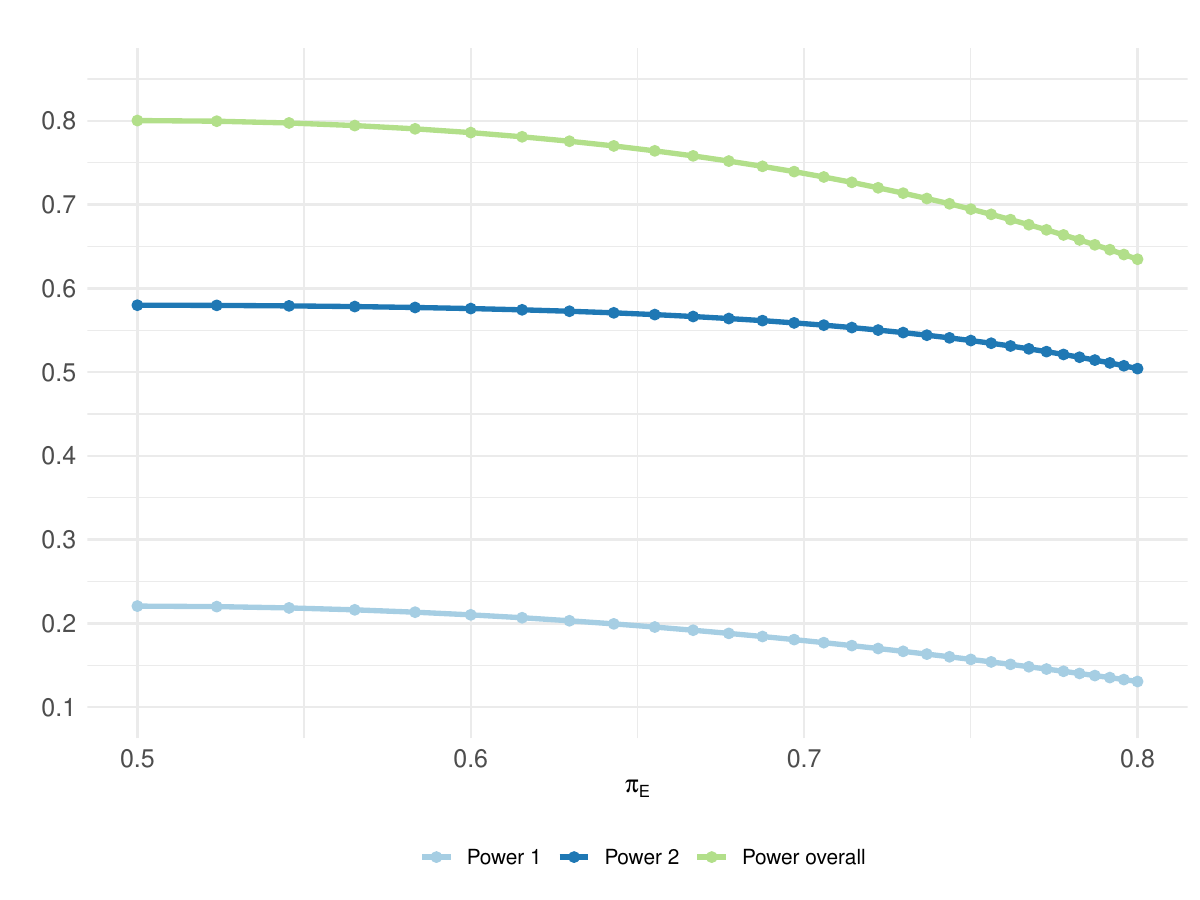} 
\caption{Theoretical loss of power (resulting from equations 6 and 7 of the main Manuscript) for the primary endpoint analysis, corresponding to an Information Fractions of 50\% and allocation proportions $\pi_E$, from 0.5 to 0.8.}
\label{fig:5}
\end{figure}

\begin{table}[ht]
\centering
\caption{Overall power using exponentially distributed outcomes with fixed allocation proportion and varying information fraction at interim analysis.}
\label{tab4}
\begin{tabular*}{\textwidth}{@{\extracolsep{\fill}} >{\centering\arraybackslash}p{3cm} *{4}{>{\centering\arraybackslash}p{2.5cm}}}
\toprule
\textbf{Allocation} & \multicolumn{4}{c}{\textbf{Information Fraction}} \\
\textbf{Fixed Proportion} & 0.2 & 0.3 & 0.4 & 0.5 \\
\midrule
0.50 & 0.80 & 0.80 & 0.80 & 0.80 \\
0.55 & 0.79 & 0.79 & 0.79 & 0.79 \\
0.60 & 0.79 & 0.79 & 0.78 & 0.78 \\
0.65 & 0.76 & 0.76 & 0.76 & 0.75 \\
0.70 & 0.74 & 0.74 & 0.74 & 0.73 \\
0.75 & 0.69 & 0.69 & 0.69 & 0.70 \\
0.80 & 0.64 & 0.63 & 0.63 & 0.63 \\
\bottomrule
\end{tabular*}
\end{table}

\clearpage

\subsection*{A Practical Guidance for Customizing the SAFER Design}

This section provides a structured and practical roadmap to assist investigators in tailoring the SAFER (Safety-Aware Flexible Elastic Randomisation) framework to their own clinical setting. The SAFER design is modular and can be adapted across different endpoint types, distributional assumptions, monitoring schemes, and inferential paradigms. The steps below outline how such customization can be implemented in practice.

\subsubsection*{S1. Definition of the Clinical Estimands}

The first step consists of clearly specifying the clinical estimands:

\begin{itemize}
    \item \textbf{Efficacy estimand:} e.g., hazard ratio, restricted mean survival time (RMST), survival probability at a fixed time point, or difference in means.
    \item \textbf{Safety estimand:} e.g., time to dose reduction/discontinuation, incidence of grade $\geq 3$ adverse events, recurrent toxicity rate, or composite safety measures.
    \item \textbf{Handling of intercurrent events:} specify whether a treatment policy, composite, hypothetical, or principal stratum strategy is adopted, in accordance with the estimands framework.
\end{itemize}

The SAFER design does not impose a specific estimand; rather, it requires consistency between the estimand definition and the statistical model used for estimation.

\subsubsection*{S2. Selection of Statistical Models}

\paragraph{Efficacy endpoint.}
Investigators may select:

\begin{itemize}
    \item Semi-parametric Cox proportional hazards models (default robust option),
    \item Parametric survival models (Weibull, Gompertz, log-normal),
    \item Flexible parametric models (e.g., spline-based hazards),
    \item Non-survival models if appropriate (e.g., binary or continuous outcomes).
\end{itemize}

If a parametric model is chosen for planning purposes, sample size and event calculations should be derived under the selected hazard structure. Inference may remain Cox-based to ensure robustness under mild model misspecification.

\paragraph{Safety endpoint.}
Possible modeling strategies include:

\begin{itemize}
    \item Time-to-event models (exponential, Weibull),
    \item Recurrent event models,
    \item Binary or ordinal regression models,
    \item Continuous toxicity burden measures.
\end{itemize}

The chosen safety model directly informs the computation of the target allocation proportion.

\subsubsection*{S3. Derivation of the Target Allocation}

The SAFER framework requires specification of a target allocation $\hat{\pi}_E$. 

\begin{itemize}
    \item Under exponential assumptions, Neyman allocation proportional to the mean time-to-event coincides with variance-based allocation.
    \item Under more general distributions (e.g., Weibull), allocation should be derived from estimated standard deviations.
    \item Alternatively, investigators may adopt a utility-based or ethical weighting approach.
\end{itemize}

The allocation rule should reflect the desired balance between statistical efficiency and ethical considerations.

\subsubsection*{S4. Calibration of the SAFER Function}

The SAFER allocation is defined as a regularized transformation of the target allocation.

Customization involves:

\begin{itemize}
    \item Selecting lower and upper allocation bounds (not necessarily 0.5 and 1),
    \item Choosing the elastic parameter $\eta$ to control aggressiveness of adaptation,
    \item Defining the transformation of efficacy evidence into the weight parameter $\hat{\Phi}$.
\end{itemize}

Typical choices:
\begin{itemize}
    \item $\eta = 1$: linear adjustment,
    \item $\eta > 1$: conservative early adaptation with more aggressive late shifts,
    \item Higher lower-bound values for conservative non-inferiority settings.
\end{itemize}

Calibration should be guided by simulation studies.

\subsubsection*{S5. Customization of the Group-Sequential Monitoring}

The SAFER design can be embedded within a flexible group-sequential framework.

Customization includes:

\begin{itemize}
    \item Number of interim analyses ($k \geq 1$),
    \item Calendar-driven or event-driven analyses,
    \item Choice of $\alpha$-spending function (O'Brien--Fleming, Pocock, Lan--DeMets),
    \item Optional $\beta$-spending for futility stopping,
    \item Conditional or predictive power stopping rules,
    \item Sample size re-estimation procedures.
\end{itemize}

The allocation updating mechanism operates independently of the monitoring schedule, preserving inferential validity.

\subsubsection*{S6. Customization of Allocation Update Timing}

Allocation updates need not coincide with interim efficacy analyses. Possible update schemes include:

\begin{itemize}
    \item Calendar-based (e.g., monthly, quarterly),
    \item Enrollment-based (e.g., every fixed number of patients),
    \item Event-driven (after a fixed number of safety events),
    \item Adaptive frequency schedules.
\end{itemize}

A burn-in period with equal allocation and/or a minimum information threshold is recommended to stabilize early estimates.

\subsubsection*{S7. Bayesian Counterpart}

A Bayesian implementation of SAFER can be naturally defined:

\begin{itemize}
    \item $\hat{\pi}_E$ derived using Thompson Sampling or posterior expected utility,
    \item $\hat{\Phi}$ replaced by the posterior probability that the treatment effect exceeds the non-inferiority margin,
    \item Interim monitoring based on posterior probability or predictive probability thresholds.
\end{itemize}

This preserves the adaptive structure while incorporating prior information and continuous belief updating.

\subsubsection*{S8. Simulation-Based Calibration}

Before implementation, extensive simulation studies should be conducted under realistic clinical scenarios to evaluate:

\begin{itemize}
    \item Type-I error control,
    \item Power,
    \item Allocation proportions over time,
    \item Adverse event rates,
    \item Robustness to model misspecification,
    \item Sensitivity to safety–efficacy association strength.
\end{itemize}

Simulation results should guide final choices of $\eta$, allocation bounds, update frequency, and monitoring strategy.

\subsubsection*{S9. Summary of Practical Customization Steps}

To adapt SAFER to a new setting, investigators should:

\begin{enumerate}
    \item Specify the clinical estimands for efficacy and safety.
    \item Select appropriate statistical models.
    \item Derive the variance-based or utility-based target allocation.
    \item Calibrate the SAFER function (bounds and $\eta$).
    \item Define group-sequential monitoring parameters.
    \item Specify allocation update timing.
    \item Conduct simulation studies to verify operating characteristics.
\end{enumerate}

The SAFER framework is therefore not tied to a specific distribution, monitoring scheme, or inferential paradigm. Its modular structure allows tailoring to different clinical contexts while preserving its core principle: balancing patient-centred safety adaptation with rigorous statistical control.

\end{document}